\documentclass[letterpaper]{article}

\usepackage{natbib,alifeconf}  
\usepackage{amsmath}
\usepackage{amssymb}
\usepackage{physics}
\usepackage{balance}
\usepackage{tabularx} 
\usepackage{amsfonts}
\usepackage{url,hyperref,cleveref}
\usepackage{booktabs}
\usepackage{booktabs}
\usepackage{lipsum}

\newcommand\blfootnote[1]{%
  \begingroup
  \renewcommand\thefootnote{}\footnote{#1}%
  \addtocounter{footnote}{-1}%
  \endgroup
}

\title{Role Differentiation in a Coupled Resource Ecology under Multi-Level Selection} 

\author{
    Siddharth Chaturvedi$^{1}$,
    Ahmed El-Gazzar$^{1}$, \and
    Marcel van Gerven$^{1}$ \\
    \mbox{}\\
    $^1$Department of Machine Learning and Neural Computing, Donders Institute for Brain, Cognition and Behaviour, Radboud University \\
    siddharth.chaturvedi@donders.ru.nl
} 

\begin{document}

\maketitle

\begin{abstract}
A group of non-cooperating agents can succumb to the \emph{tragedy-of-the-commons} if all of them seek to maximize the same resource channel to improve their viability. In nature, however, groups often avoid such collapses by differentiating into distinct roles that exploit different resource channels. It remains unclear how such coordination can emerge under continual individual-level selection alone. To address this, we introduce a computational model of multi-level selection, in which group-level selection shapes a common substrate and mutation operator shared by all group members undergoing individual-level selection. We also place this process in an embodied ecology where distinct resource channels are not segregated, but coupled through the same behavioral primitives. These channels are classified as a positive-sum intake channel and a zero-sum redistribution channel. We investigate whether such a setting can give rise to role differentiation under turnover driven by birth and death. We find that in a learned ecology, both channels remain occupied at the colony level, and the collapse into a single acquisition mode is avoided. Zero-sum channel usage increases over generations despite not being directly optimized by group-level selection. Channel occupancy also fluctuates over the lifetime of a boid. Ablation studies suggest that most baseline performance is carried by the inherited behavioral basis, while the learned variation process provides a smaller but systematic improvement prior to saturation. Together, the results suggest that multi-level selection can enable groups in a common-pool setting to circumvent tragedy-of-the-commons through differentiated use of coupled channels under continual turnover.
\end{abstract}

Submission type: \textbf{Full Paper}\\

Data/Code available at: \url{https://abmax.short.gy/multi-level-selection-ecology}
\blfootnote{\textcopyright  2026 [AUTHORS' NAMES]. Published under a Creative Commons Attribution 4.0 International (CC BY 4.0) license.}

\section{Introduction}

Heterogeneity in groups of living systems, such as differentiation of roles, is a common phenomenon that is often important for their survival~\citep{rueffler2012evolution}. It can be observed at various scales of their existence, from different types of cells in an organ to different roles in a colony of organisms. Such heterogeneity can be inferred from the different functional niches that group members evolve to occupy. For instance, in animal colonies such as ants and bees~\citep{oster1978caste}, individuals may differ in how they acquire resources and contribute to group maintenance. This creates functional heterogeneity within a shared ecological system and exemplifies division of labour~\citep{rueffler2012evolution}. If group members reside in a common-pool-resource environment~\citep{levin2014public}, such heterogeneity can help them circumvent the \emph{tragedy-of-the-commons} scenario~\citep{hardin1968}. In this scenario, if all individuals seek to optimize a resource via the same channel, the entire group may exhaust it to starvation.

However, it seems implausible that individual group members are selected by evolution for their ability to optimize a group-level objective~\citep{traulsen2008analytical, tudge2016game}. On the contrary, it is much more plausible that selection at the level of individual members is relative. Members with higher fitness get more opportunities to reproduce. Thus, competition among individuals in a common-pool-resource environment is a natural mode of interaction. On the other hand, uncovering the mechanism responsible for the emergence of coordination, such as heterogeneity or role differentiation, in such settings still remains an open question~\citep{west2015major}. At a deeper level, this question also relates to major evolutionary transitions. These include the emergence of multicellular organization from interacting unicellular units~\citep{michod2007evolution}, or scaling of individual-level goals to group-level objectives in living systems~\citep{pio2023scaling}.

One source of diversity can be attributed to the fact that in many natural settings, selection does not strictly follow the \emph{survival-of-the-fittest} rule but instead operates through a minimal-criteria selection~\citep{brant2017minimal}. Whereby, all individuals who are able to maintain their fitness above a certain minimal threshold remain viable for reproduction, compared to only the elite individuals getting the opportunities. Computational models of such a selection process have already shown the emergence of different foraging modes in populations of similar foragers~\citep{aubert2015hunger, hamon2023eco}. However, since the foraging channel for all the foragers in those models was the same, the overall population behavior still remained vulnerable to the tragedy-of-the-commons.

At the group-level, from a Darwinian perspective, it can be argued that groups of living systems that can resolve the tragedy-of-the-commons scenario more successfully are better candidates for natural selection than those that can not~\citep {traulsen2008analytical}. The interplay between individual-level selection and group-level selection is often formulated as a multi-level selection process~\citep{shelton2020group}. Computational models of such interplay have demonstrated the evolution of altruistic individuals even when altruism was associated with lower fitness at the individual-selection level~\citep{doekes2024multiscale}.

In this work, we present a computational model of a multi-level selection process where we explore a specific type of coupling between the two levels of selection. Namely, the group-level selection acts on a shared controller substrate, together with a mutation-operator, both of which are expressed during individual-level minimal-criteria selection. The shared controller substrate provides the inherited dynamical regime within which individual-level selection unfolds, while the mutation-operator further guides the local generation of variation during reproduction. The aim of the model is to investigate if such a coupling can give rise to role differentiation, via which the group does not succumb to the tragedy-of-the-commons. More specifically, the goal is not merely to obtain role differentiation once, but to investigate whether such differentiation can be continually regenerated under ongoing turnover fueled by the birth and death of agents. The hypothesis is based on the fact that, firstly, in natural multicellular development, multiple functional cellular states can arise from the same genotype through a common underlying dynamical basis~\citep{sole2024open}. Secondly, mutation-operators may not only favor the short-term fitness of organisms but also the long-term benefits of future mutations~\citep{ferrare2024evolution}. In our case, these long-term benefits are evaluated through the group-level fitness.

We use an active-particle-based boid model~\citep{Reynolds87}, where agents can acquire resources from two sources. They can either slow down relative to the global average and \emph{graze} resources in direct proportion to how slow they move, or they can exchange resources among themselves during close encounters, where resources flow from agents in relative abundance to agents in relative scarcity in proportion to the difference. Such a construction is ecologically meaningful because the same behavioral degrees of freedom can simultaneously shape resource gain, redistribution, and loss, and ecological theory often explains coexistence in common-pool settings through such coupled trade-offs rather than through isolated resource options~\citep{werner1993ecological, kneitel2004trade, levin2014public}. The goal of the multi-level selection then becomes to discover a shared controller substrate and a mutation-operator that favors producing agents that show both resource-acquisition modes co-existing with each other under constant turnover of agents. In the following section, we describe the model in detail, after which we present the main results and conclusions of the work.

\section{Model}
\begin{figure}
    \centering
    \includegraphics[width=1.0\linewidth]{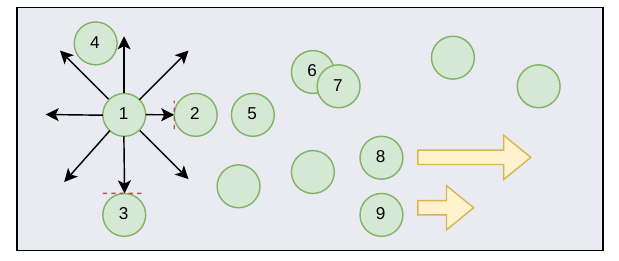}
    \caption{
        \textbf{Simulation sample scenario.} Boids $2$ and $3$ are detected by the rays emitted from boid $1$, while boid $4$ lies in the blind-spot of $1$. Boid $2$ occludes boid $5$ from $1$. Boids $6$ and $7$ overlap, thus exchange resources. Boid $8$ moves faster, thus incurs more metabolic-cost, while boid $9$ is slower, thus gains from grazing.
    }
    \label{fig:model}
\end{figure}

The model used in this work consists of circular disk-shaped agents of equal diameter, called boids, moving in a two-dimensional Cartesian plane. The boids can pass through each other without a hard body collision. In each rollout, at most $N_{\max}$ boids can exist simultaneously, while only a subset $N$ of them may be active at a given time. Active boids move, sense, exchange resources, reproduce, and die (by becoming inactive). Inactive boids occupy dormant slots outside the arena until they are reactivated by reproduction. The boids move using an acceleration control model based on a stochastic double integrator. Each rollout always starts with $N_{\min}$ boids that are immortal and help kick-start the process of population development. All dynamical updates are integrated using Euler's forward method, with time $t \in \{0,\dots, T\}$ advanced in steps of size $\dd t$. The model updates are further organized into the following submodels. Figure~\ref{fig:model} summarizes all the possible interactions a boid can experience in the model. 

\subsection{Position model}
The center of a boid $i$ in the model is given by $\mathbf{q}_i(t)\in \mathbb{R}^2$ and its orientation is given by $\theta_i(t)\in(-\pi,\pi]$. After dropping the boid index $i$ for notational convenience, boid positions are updated as
\begin{equation}\label{eq:position}
    \dot{\mathbf{q}}(t) = \mathbf{v}(t), \,\,
    \dot{\mathbf{v}}(t) = s(t)\!\begin{bmatrix}\cos\theta(t)\\ \sin\theta(t)\end{bmatrix} - \lambda\mathbf{v}(t)
\end{equation}
and the orientations are updated as
\begin{equation}\label{eq:orientation}
        \dot \theta(t) =  \omega(t), \,\, \dot \omega(t) = u(t) -\lambda\omega(t)
\end{equation}
where $\mathbf{v}(t)\in\mathbb{R}^2$ is its velocity along the Cartesian axes and $\omega(t)\in\mathbb{R}$ is its angular velocity. The plane's damping coefficient is given by a positive constant $\lambda$. The quantities $s(t)\in\mathbb{R}$ and $u(t) \in\mathbb{R}$ are the translational and rotational control outputs of a boid's acceleration controller, respectively, and are given by
\begin{align}\label{eq:speed}
    s(t) &= \frac{d_b}{2}\tanh\left(a^{(s)}(t)\right)\cdot \left(1 + \epsilon\xi^{(s)}(t)\right)\\
    u(t) &= \tanh\left(a^{(u)}(t)\right)\cdot\left(1 + \epsilon\xi^{(u)}(t)\right)
\end{align}
where $\epsilon$ is a positive scalar by which Gaussian noise terms $\xi^{(s)}(t)$ and $\xi^{(u)}(t)$ are scaled. Based on active-particle modeling conventions~\citep{fily2012athermal}, the common radius of boids $0.5 d_b$ is used to scale the translational motion with respect to rotational motion. The quantities $a^{(s)}(t), a^{(u)}(t) \in \mathbb{R}$ are the readouts of a boid's acceleration controller. Lastly, the values of $\|\mathbf{q}(t)\|$, $\|\mathbf{v}(t)\|$ and $|\omega(t)|$ are always bounded by $q_{\max}$, $v_{\max}$ and $\omega_{\max}$ respectively, for all the boids.

\subsection{Resource model}
A boid's internal resource depot is given by $e(t)$. The boids can gain and lose resources through three distinct mechanisms. Firstly, when the movement $m_i(t)$ of a boid $i$, given by
\begin{equation}\label{eq:movement}
    \dot m_i(t) = \tau^{(m)}\left(||\mathbf{v}_i(t)|| + |\omega_i(t)| - m_i(t)\right)
\end{equation}
falls below the average movement $m_N(t)$ of all the boids, given by
\begin{equation}\label{eq:avg_movement}
    m_N(t) = \frac{1}{N}\sum_{j\in N}m_j(t),
\end{equation}
 boid $i$ obtains resources via the \emph{grazing} channel given by
\begin{equation}
    e_i^{(g)}(t) = k^{(g)}(m_N(t) - m_i(t)).
\end{equation}
Here, $\tau^{(m)}$ is a time constant by which the norm on boids' speeds accumulates into its movement value, and $k^{(g)}$ is a positive scaling constant.
Secondly, when two boids $i$ and $j$ pass through each other, a resource value proportional to the difference of their internal resource depot is transferred from the boid having more resources to the boid having less resources. Thus, for each boid $i$ the net resources gained or lost via the resources \emph{exchange} channel $e_i^{(e)}$ is given by
\begin{equation}
    e_i^{(e)}(t) = k^{(e)}\sum_{j\in N}(e_j(t) - e_i(t))\delta_{ij}(t)
\end{equation}
 where $\delta_{ij}(t)$ is a boolean function that is \emph{True} when agents $i$ and $j$ overlap and \emph{False} otherwise. Two boids overlap when the Euclidean distance between their centers is less than their common diameter $d_b$. Further, $k^{(e)}$ is a positive scaling constant signifying the resource transfer rate. Finally, boids also lose the resource via \emph{metabolic losses}, given as
 \begin{equation}
      e^{(c)}_i(t) = k^{(c)(s)}|s_i(t)| + k^{(c)(u)}|u_i(t)| + \gamma e_i(t)
 \end{equation}
 where $k^{(c)(s)}$ and $k^{(c)(u)}$ are positive scaling constants and $\gamma$ is the leak coefficient. Thus, the change in resource value for boid $i$ is given by
 \begin{equation} \label{eq:energy_update}
     \dd e_i(t) = e_i^{(g)}(t) + e_i^{(e)}(t) - e_i^{(c)}(t).
 \end{equation}
Quantities $e_i(t)$, $e^{(g)}_i(t)$ and $e^{(e)}_i(t)$ are always clipped such that they are non-negative and less than $e_{\max}$, $e^{(g)}_{\max}$ and $e^{(e)}_{\max}$ respectively.

It is worth noting that the grazing and exchange channels are designed such that they create a behavioral tension in the system. Grazing rewards boids that move slower than the population average, thereby encouraging reduced movement. In contrast, the exchange channel depends on spatial overlap between boids, and is more likely to occur under higher relative motion and encounters. This can induce interactions where boids with lower resource values tend to move towards others, while those with higher resource values tend to avoid overlap. Further, the grazing, exchange, and metabolic loss instantiate a positive, zero, and negative-sum flow of resources into the system, respectively.

\subsection{Sensor and controller model}
At all time instances, each boid $i$ emits $r$ equiangular rays into its environment from its position $\mathbf{q}_i(t)$. Upon striking the surface of another boid, each ray $k$ gathers two channels of information given by $\left[d_k(t),\, e_k(t) \right]$. If a ray does not strike a surface, then it gathers $\left[d_{\max},\, 0.0 \right]$ as information for the controller. The rays are also subjected to occlusion in the environment (see Figure~\ref{fig:model}). The ray-surface intersections are computed via the standard ray-circle quadratic test~\citep{ericson2004real}, similar to \cite{chaturvedi2025emergence}. Thus, the net ray sensor vector is given by
\begin{equation}
    \mathbf{o}^{(r)}_i(t) = \left[d_1, e_1,\dots,d_r, e_r\right].
\end{equation}
Information gathered by all the rays is further concatenated with other boid states given by
\begin{equation}
    \mathbf{o}^{(b)}_i(t) = \left[\mathbf{v}_i, \omega_i, m_i, m_N, e_i, e_i^{(g)}, e_i^{(e)}, e_i^{(c)}, w_i\right]
\end{equation}
where $\mathbf{v}_i(t)$, $\omega_i(t)$ are the boid velocity components (Equation~\ref{eq:position} and~\ref{eq:orientation}), $m_i(t)$ and $m_N(t)$ are the boid movement and the average movement of all the boids in the environment respectively (Equation~\ref{eq:movement} and ~\ref{eq:avg_movement}), $e_i(t)$, $e_i^{(g)}(t)$, $e_i^{(e)}(t)$ and, $e_i^{(c)}(t)$ represent the value of resource depot, resource gained or lost via the grazing, exchange and metabolic loss channels respectively (Equation~\ref{eq:energy_update}), and $w_i(t)$ is the number of boids with which $i$ overlaps. The final observation vector fed into a boid's controller is given by $\mathbf{o}_i = \bigl[\mathbf{o}_i^{(r)}(t), \mathbf{o}_i^{(b)}(t)\bigr]^\top \in \mathbb{R}^{(2r +10)}$.

Next, each boid is controlled by a continuous-time recurrent neural network (CTRNN) with $n$ hidden units. More specifically, the CTRNN is based on a modern interpretation of the Wilson-Cowan model of rate-based activity in neuronal populations~\citep{sussillo2014neural} and is given by
\begin{equation}\label{eq:CTRNN}
    \dot{\mathbf{z}}_i = \mathbf{\tau^{(z)}}\odot(J_i\cdot(\sigma(\mathbf{z}_i(t) +\mathbf{b})) +E\mathbf{o}_i(t)-\mathbf{z}_i(t))
\end{equation}
where for boid $i$, $\mathbf{z}_i(t)\in \mathbb{R}^n$ is the activation vector of the hidden units, $\mathbf{\tau}^{(z)}\in \mathbb{R}^n$ is a vector of time constants for each unit, $J_i \in \mathbb{R}^{n\times n}$ is the recurrent inter-connectivity matrix, $\mathbf{b} \in \mathbb{R}^n$ is a vector of biases, $\mathbf{o}_i(t)\in \mathbb{R}^{2r+10}$ is the observation vector of the controller and, $E\in \mathbb{R}^{n\times (2r+10)}$ is the observation-scaling matrix. Further, $\sigma(\cdot)$ represents the sigmoid activation function, and the operator $\odot$ represents an element-wise multiplication of two vectors. The outputs of the controller to the position model are read out as
\begin{equation}\label{eq:controller_readout}
    \begin{bmatrix}
        a_i^{(s)}(t), 
        a_i^{(u)}(t)
    \end{bmatrix}^\top = \tanh(D\cdot \mathbf{z}_i(t))
\end{equation}
where $D\in\mathbb{R}^{2\times n}$ is a readout scaling matrix. It is further worth noting that the parameter set $\phi=\{\mathbf{\tau^{(z)}}, \mathbf{b}, E, D\}$ is common for all the boids in a population, and they form the shared controller substrate tuned by the group-level selection. In contrast, the inter-connectivity matrix $J_i$ is unique for a boid $i$ and is acted upon by the mutation-operator during the individual-level selection.
\begin{figure}
    \centering
    \includegraphics[width=1.1\linewidth]{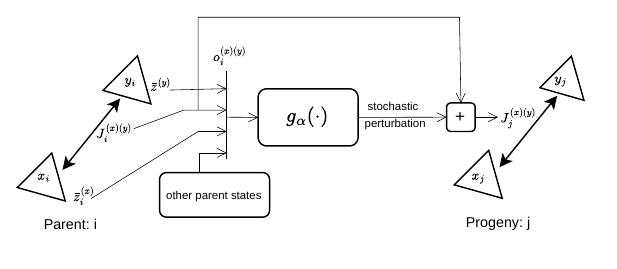}
    \caption{\textbf{Mutation process.} The mutation-operator $g_\alpha(\cdot)$ acts on each element $J_i^{(x)(y)}$ of the parent boid's $J_i$ using the input $\mathbf{o}_i^{(x)(y)}$. The resulting perturbation, together with stochastic noise, is added to $J_i^{(x)(y)}$ to produce the corresponding element $J_j^{(x)(y)}$ of the progeny boid's $J_j$.
     }
    \label{fig:mutation}
\end{figure}

\subsection{Individual-level selection}
The process of individual-level selection is modeled based on the minimal-criteria selection~\citep{brant2017minimal, hamon2023eco}, where agents that are able to maintain a viability criterion over a minimal threshold for a certain minimum amount of time are viable for asexual reproduction, while the agents that allow a drop in their viability levels below a threshold are eliminated. More specifically, in our model, the two processes take place in the following manner.

\textbf{\emph{Birth:}} When a boid $i$ is able to maintain its resource depot $e_i(t)$ above a certain threshold $e_{\text{birth}}$ for an uninterrupted time interval of $t_{\text{birth}}$ ($e_i(t') \geq e_{\text{birth}} \forall t'\in(t-t_{\text{birth}},t)$), it asexually produces a progeny boid $j$ with stochastic mutations. The mutations are produced by a mutation-operator $g(\cdot)$ modeled using a small multi-layered perceptron (MLP) with $l$ hidden layers and $h$ units per layer. At the time of reproduction $t$, the mutation-operator exclusively acts on individual components of the inter-connectivity matrix $J_i$ of the parent's controller to produce the inter-connectivity matrix $J_j$ of the progeny's controller, such that
\begin{equation}
    J^{(x)(y)}_j = J^{(x)(y)}_i + g(\mathbf{o}^{{(x)(y)}}_i(t))(1+\eta\xi^{{(x)(y)}}(t))
\end{equation}
where Gaussian noise $\xi^{{(x)(y)}}(t)\in\mathbb{R}$ is scaled by a positive constant $\eta$ and $\mathbf{o}_i^{(x)(y)}(t)\in \mathbb{R}^8$ is the input to the mutation-operator $g$ such that $g$ produces a mutation perturbation for each element $J^{(x)(y)}_i$ of $J_i$, and it is given by
\begin{equation}
    \mathbf{o}_i^{(x)(y)} = \left[\bar{z}_i^x,\, \bar{z}_i^y,\, J^{(x)(y)}_{i},\, \bar{e}_i,\, \bar{e}^{(g)}_i,\, \bar{e}^{(e)}_i,\, \bar{e}^{(c)}_i,\, m_i \right]
\end{equation}
where $x,y \in \{1,\dots,n\}$ represent the indexing of the matrix rows and columns, respectively. Further, from a loose Hebbian perspective~\citep{hebb2005organization}, $\bar{z}_i^x\in \mathbb{R}$ and $\bar{z}_i^y\in \mathbb{R}$ can be inferred as a moving average of the neuronal activity of the output neuron $z^{(x)}_i(t)\in \mathbf{z}_i(t)$ and the input neuron $z^{(y)}_i(t)\in \mathbf{z}_i(t)$ respectively, with $J^{(x)(y)}_{i}$ being the inter-neuron connection or \emph{synapse} between them. The moving average filter for a neuron's activity is given as
\begin{equation}\label{eq:moving average}
    \dot {\bar{z}}(t) = \tau^{(\bar{z})}(z(t) - \bar{z}(t)).
\end{equation} Similarly, $\bar{e}_i(t)$, $\bar{e}^{(g)}_i(t)$, $\bar{e}^{(e)}_i(t)$ and, $\bar{e}^{(c)}_i(t)$ are moving averages of the boid's current resource depot value $e_i(t)$ and the amount of resources recently gained or lost via the grazing ${e}^{(g)}_i(t)$, exchange ${e}^{(e)}_i(t)$, and ${e}^{(c)}_i(t)$ metabolic cost channels respectively. These moving averages follow the same dynamics as given by Equation~\ref{eq:moving average} on their respective variables with a common time constant $\tau^{(\bar{e})}$. It can be noted that by design, the mutation-operator operates on the moving averages of certain boid states instead of their instantaneous values in order to use the average state of the parent over a longer time horizon rather than over the single time instant of birth. The mutation-operator is shared by all the boids in a population and is summarized in Figure~\ref{fig:mutation}. Further, the progeny is spawned randomly within a radius of $d_{\text{spawn}}$ near its parent and inherits half the resource depot from its parent ($e_j(t) := 0.5e_i(t)$). Inspired by the process of mitosis in living cells, during birth, the resource depot of the parent is also halved ($e_i(t) := 0.5e_i(t)$). This reduction in the parent's resource depot introduces a natural delay before which the parent can reproduce again, and prevents population explosions. Finally, the progeny is initialized with the same controller substrate $\phi$ as that of its parent and all the other boids in its population.

\begin{figure}
    \centering
    \includegraphics[width=1.0\linewidth]{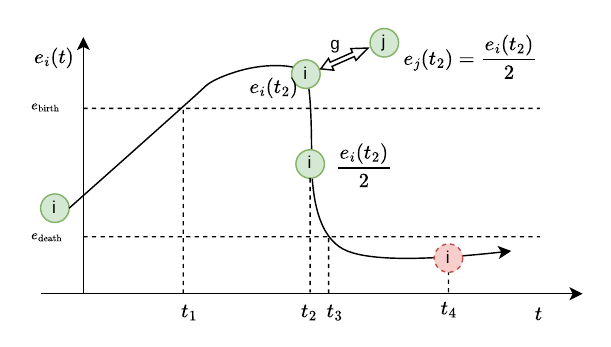}
    \caption{\textbf{Birth-death process.} When boid $i$ spends time $t_2-t_1 > t_{\text{birth}}$ above the resource threshold for birth ($e_{\text{birth}}$), it produces a progeny $j$ with half the resource at that step $e(t_2)/2$. In the process, its own resource depot is halved. When boid $i$ spends time $t_4-t_3>t_{\text{death}}$ below the resource threshold for death $e_{\text{death}}$, it is eliminated from the simulation.
    }
    \label{fig:birth_death}
\end{figure}

\textbf{\emph{Death}}: As soon as an active boid's internal resource depot $e_i(t)$ drops below a threshold $e_{\text{death}}$ for an uninterrupted time interval of $t_{\text{death}}$ ($e_i(t') \leq e_{\text{death}} \forall t'\in(t-t_{\text{death}},t)$), the boid is eliminated from the population. Further, $e_{\text{death}}$ is set such that $e_{\text{death}}<e_{\text{birth}}$ and there are always $N_{\min}$ boids in the beginning of each rollout that are labeled as \emph{immortals} and cannot be eliminated. This is done to prevent a complete population crash and kick-start the process of population regrowth during a rollout. When a boid's age $a_i(t)$ exceeds a certain threshold $a^{\min}_\text{old}$, the boid $i$ is susceptible to a random elimination based on Bernoulli probability sampling. The probability of elimination rises linearly from $0.0$ at $a^{\min}_\text{old}$ to $1.0$ at $a^{\max}_\text{old}$. Figure~\ref{fig:birth_death} depicts an example of the birth and death scenario.

\subsection{Group-level selection}
The controller substrate represented by $\phi=\{\mathbf{\tau^{(z)}}, \mathbf{b}, E, D\}$ and the parameters of the multi-layer perceptron ($\alpha$) representing the mutation-operator $g_\alpha(\cdot)$ are common to all the boids in a population or a group and therefore take part in the group-level selection process.
The fitness of a group is defined using two criteria. Firstly, a group of boids that can maximize the positive intake of resources and minimize its negative loss at a group-level is assigned a higher fitness. In our setup, the grazing channel represents the positive influx of the resource and the metabolic cost represents the negative drain thus the resource fitness $f^{(e)}$ is given as
\begin{equation}\label{eq:resource_fitness}
    f^{(e)} = \sum_{i}\left(\sum_{t}(e_i^{(g)}(t) - e_i^{(c)}(t))\right)
\end{equation}
where the inner sum runs over all rollout time steps and the outer sum over all boids. It is important to note that boids that were active during the rollout but were eliminated before $T$ also contribute towards this fitness. Also, since the exchange channel gives rise to a zero-sum game in the interacting boids of a group it can be omitted from Equation~\ref{eq:resource_fitness}. 
Secondly, a group with higher age-mass is assigned more fitness. The age-mass fitness $f^{(a)}$ of a group is given by
\begin{equation}\label{eq:fitness_age_mass}
    f^{(a)} = \sum_{i}\left(\sum_{t}(\delta^{(a)}_i(t)a_i(t))\right)
\end{equation}
where $\delta^{(a)}_i(t)$ is a boolean flag that is $1$ if the boid $i$ is active and $0$ otherwise and $a_i(t)$ is the age of boid $i$ in the rollout, which begins at $a_i=0$ when the boid is born and is incremented at each time step by $1$ until the boid is eliminated. The age-mass term is included to favor groups that sustain viable individuals over time, rather than relying only on short-lived agents that contribute transiently to the resource balance.
The final fitness of a group $k$ thus becomes 
\begin{equation}\label{eq:net_fitness}
    f_k = f_k^{(e)} + \mu f_k^{(a)}
\end{equation}    
with $\mu$ being a scaling factor. The aim of the group-level selection is to select a group with optimal parameter set $\{\phi^\star,\alpha^\star\}$ such that
\begin{equation}
    \{\phi^\star,\alpha^\star\} = \arg\max_{\phi_k,\alpha_k}f_k(\phi_k,\alpha_k).
\end{equation}
For this purpose we used an evolutionary strategy, namely the covariance matrix adaptation evolutionary strategy (CMA-ES)~\citep{hansen2016cma}. In practice, for each generation, we collect all the parameters of the set $\{\phi_k,\alpha_k\}$ in a vector and subject it for optimization by the CMA-ES algorithm for $M$ groups serving as the population size. Also, each parameter vector is first evaluated across $S$ different scenarios, corresponding to different initial conditions generated from different random seeds, and the final fitness is taken as the average across these $S$ scenario-wise fitness values. The \texttt{Evosax} library~\citep{lange2023evosax} was used to realize the CMA-ES algorithm in practice.

\subsection{Simulation details}
The model can be simulated in two modes, namely, training mode and inference mode. The numerical values of model parameters and initial values of the state in the training mode are listed in Table~\ref{tab:model_training_values}. Unless otherwise stated these values also serve as the default values of the model in the inference mode. The model was realized in \texttt{ABMax}~\citep{chaturvedi2025abmax} -- an agent-based modeling framework written in Python \texttt{JAX}~\citep{jax2018github}. The rank-match algorithm from \texttt{ABMax} was used to model individual-level selection efficiently. This allowed batched updates across boid groups, even when the population in each group changed dynamically and independently.

\begin{table}[!t]
    \centering
    \begin{small}
    \begin{tabular}{lll}
    \toprule
    \text{Notation} & \text{Description} & \text{Value} \\ \midrule
    $\dd {t}$ & Step size for simulations & $0.1$ \\
    $T$ & Total time steps in a rollout & $4000$ \\
    $N_{\max}$ & Maximum number of boids in a group & $50$ \\
    $N_{\min}$ & Number of immortal boids in a group & $5$ \\
    $\lambda$ & Damping coefficient of the plane & $0.3$ \\
    $d_b$ & Diameter of boids & $20$ \\
    $q_{N_{\min}}(0)$ & Initial immortal boid positions & $U(\pm50)$ \\
    $q_{\max}$ & Cartesian plane limit & $10{,}000$ \\
    $\theta(0)$ & Initial boid orientations & $U(\pm\pi)$ \\
    $v_{\max}$ & Translational speed limit & $20.0$ \\
    $\omega_{\max}$ & Rotational speed limit & $\pi/3$ \\
    $\mathbf{v}(0)$ & Initial Cartesian velocity & $[0.0,\,0.0]^\top$\\
    $\omega(0)$ & Initial angular velocity & $0.0$ \\
    $\epsilon$ & Noise scaling (position model) & $0.1$ \\
    $\tau^{m}$ & Time constant (movement) & $0.04$ \\
    $m(0)$ & Initial movement of boids & $0.0$ \\
    $k^{(g)}$ & Scaling coefficient (grazing) & $0.1$ \\
    $k^{(e)}$ & Exchange scaling & $0.2$ \\
    $k^{(c)(s)}$ & Translational metabolic-cost scaling & $0.004$ \\
    $k^{(c)(u)}$ & Angular metabolic-cost scaling & $0.04$ \\
    $\gamma$ & Resource leak coefficient & $0.001$ \\
    $e_{\max}$ & Maximum resource in a boid & $100.0$ \\
    $e^{(g)}_{\max}$ & Maximum resource grazed in a step & $4.0$ \\
    $e^{(e)}_{\max}$ & Maximum resource exchanged in a step & $8.0$ \\
    $e_{N_{\min}}(0)$ & Initial resource (immortal boids) & $U(20,50)$ \\
    $r$ & Number of rays emitted by a boid & $11$ \\
    $d_{\max}$ & Maximum ray reach & $300.0$ \\
    $J_{N_{\min}}$ & Immortals CTRNN $J$ matrix & $U(\pm1)$\\
    $z(0)$ & Initial CTRNN hidden states & $0.0$\\
    $n$& Number of CTRNN hidden states & $40$\\
    $e_{\text{birth}}$ & Resource threshold (reproduction) & $20.0$\\
    $t_{\text{birth}}$ & Time step threshold (reproduction) & $40$\\
    $\bar{s}(0)$ &Initial value moving average (any state $s$) & $s(0)$\\
    ${\tau}^{\bar{z}}$ &Time constant (CTRNN moving average) & $0.04$\\
    ${\tau}^{\bar{e}}$ &Time constant resource states & $0.04$\\
    $\eta$ & Noise scaling (mutation) & $0.05$ \\
    $l$ & Hidden layers (mutation-operator MLP) & $2$ \\
    $h$ & Units per layer (mutation-operator MLP) & $16$ \\
    $e_{\text{death}}$ & Resource threshold (death) & $2.0$\\
    $t_{\text{death}}$ & Time step threshold (death) & $40$\\
    $a^{\min}_{\text{old}}$ & Old-age onset time step & $400$\\
    $a^{\max}_{\text{old}}$ & Old-age conclusion time step & $600$\\
    $\mu$ & Age fitness scaling & $5.0\times10^{-8}$ \\
    $M$ & Number of groups (CMA-ES population)& $50$ \\
    $S$ & Number of scenarios & $2$\\
    - & CMA-ES distribution elite ratio & $0.3$\\
    - & CMA-ES initial step size & $0.1$\\
    - & Number of generations & $2000$\\
    \bottomrule
    \end{tabular}
    \caption{Parameters used for simulations}
    \label{tab:model_training_values}
    \end{small}
\end{table}

\section{Results}
In this section we present the main results of the optimization process and some observations after simulating the model in an inference mode for longer time horizons. We also perform a minimal ablation analysis to check how different modeling components contribute towards the results.
\begin{figure}
    \centering
    \includegraphics[width=0.9\linewidth]{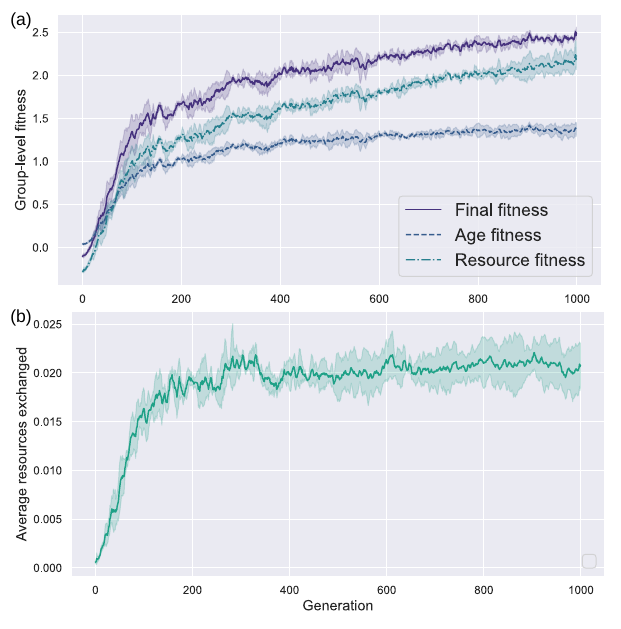}
    \caption{\textbf{Group-level fitness trends: (a)} The resource fitness, age-mass fitness, and net fitness rise and saturate across generations. \textbf{(b)} The average positive resources received via exchange by boids also increases with generations. Trends in \textbf{(a)} and \textbf{(b)} are averaged across $3$ training seeds and smoothed using a filter of $5$ bins.
    }
    \label{fig:fitness_curve}
\end{figure}
\subsection{Group-level optimization}
The fitness curve of the group-level selection optimization is shown in Figure~\ref{fig:fitness_curve}(a). It can be seen that both components of the net group-level fitness in Equation~\ref{eq:net_fitness}, namely net resource intake and age-mass, show an increasing trend across generations. We further define the average positive resources received via exchange by boids in generation $g$ as
\begin{equation}\label{eq:posiive_exchange}
e^{(e+)}_{\text{mean}}(g)
=
\frac{1}{Z}
\sum_{k=1}^{M}
\sum_{s=1}^{S}
\sum_{i=1}^{N_{\max}}
\sum_{t=1}^{T}
\max\!\left(e^{(e)}_{i,k,s,g}(t),\,0\right)
\end{equation}
where $k$ indexes groups in the CMA-ES population, $s$ indexes rollout scenarios, $g$ is the generation index, and $Z = M S N_{\max} T$. As shown in Figure~\ref{fig:fitness_curve}(b), this quantity also increases across generations, despite not being directly optimized by the group-level fitness. This suggests that the group-level selection can indirectly enhance the use of the zero-sum exchange channel while acting only on net positive-sum intake together with age-mass.

\subsection{Role differentiation}
\begin{figure*}[!t]
    \centering
    \includegraphics[width=\textwidth, height=0.78\textheight, keepaspectratio]{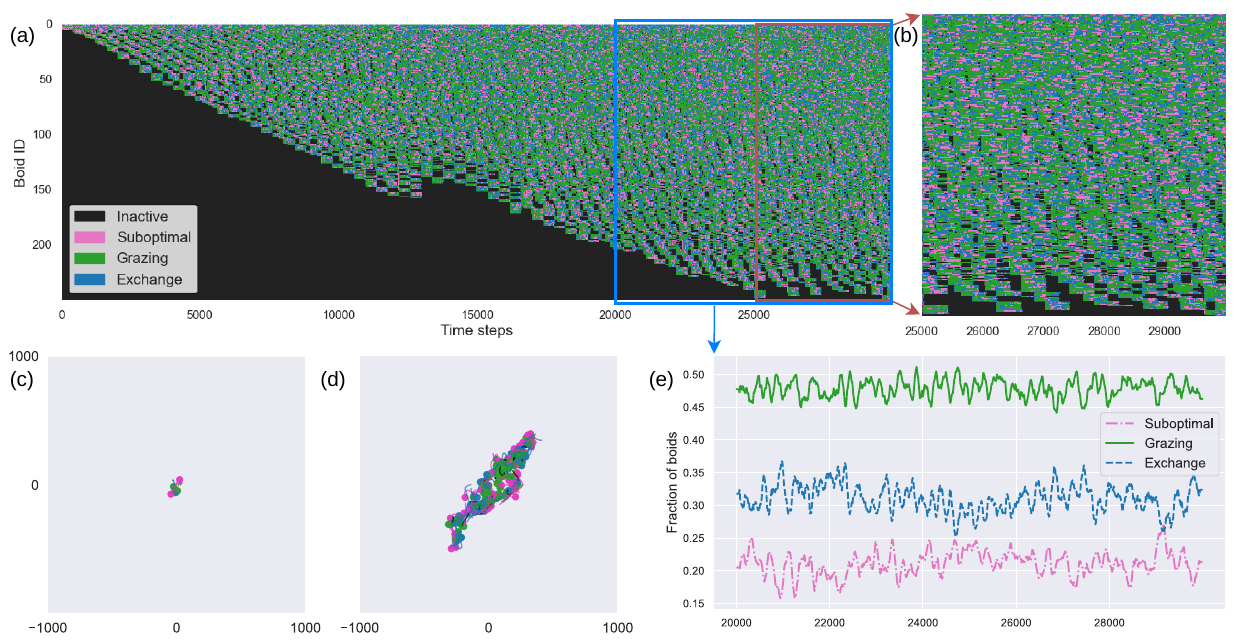}
    \caption{ \textbf{Inference analysis: (a)} Different roles observed in boids during a long inference rollout. \textbf{(b)} A zoomed-in view of the role heat map for the same rollout between the time steps $25{,}000$ and $30{,}000$. \textbf{(c)} Cartesian plane boid trajectories in the first $100$ time steps of the rollout. \textbf{(d)} Cartesian plane boid trajectories between time steps $20{,}000$ and $20{,}100$. \textbf{(e)} Proportion of active boids in different roles between time steps $20{,}000$ and $30{,}000$.
    }
    \label{fig:heatmap}
\end{figure*}

In an inference mode, we run the optimized model for $T=30{,}000$ time steps with $N_{\max}$ extended to $250$ boids. Figures~\ref{fig:heatmap}(c) and (d) depict the trajectories of the active boids near the beginning and near maturity of the rollout~\footnote{A video is available at: https://youtu.be/lXzLKRaL4PY} respectively. It can be seen that, near maturity, the boids remain spatially clustered, which is consistent with sustained occupation of the exchange channel.

Next, we define $3$ roles that a boid can occupy based on the dominant channel of its resource balance. Thus, for a boid $i$, if the moving average of the positive resource intake via the exchange channel, ${\bar{e}}^{(e+)}_{i}(t)$, given by
\begin{equation}
    \dot{\bar{e}}^{(e+)}_{i}(t) = \tau^{(\bar{e})}\left(\max\!\left(0, e^{(e)}_i(t)\right)-\bar{e}^{(e+)}_{i}(t)\right)
\end{equation}
exceeds the moving averages of resources gained via grazing, $\bar{e}_i^{(g)}(t)$, and resources lost via metabolic cost, $\bar{e}_i^{(c)}(t)$, then the boid is labeled to be in the \emph{exchange} role. If the grazing channel dominates the other two by the same logic, the boid is labeled to be in the \emph{grazing} role. Otherwise, the boid is labeled \emph{suboptimal}. A heat-map for the same is depicted in Figures~\ref{fig:heatmap}(a) and (b) for the rollout.

It can be seen that, firstly, the number of active boids increases over time. More importantly, heterogeneous channel occupancy is maintained not only across the boids active at a given time, but also across the lifetime of an individual boid. Further, although individual boids switch roles over time, the colony maintains approximately stable role proportions near maturity.

\subsection{Ablation analysis}
\begin{figure}
    \centering
    \includegraphics[width=0.9\linewidth]{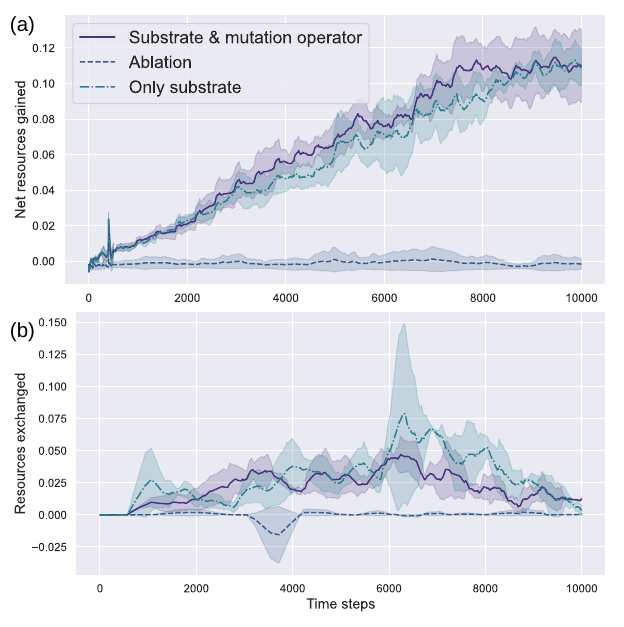}
    \caption{\textbf{Ablation analysis:} \textbf{(a)} Comparison of the average net resources gained by a group in the $3$ settings. \textbf{(b)} Comparison of the average positive resources received via exchange. Trends in \textbf{(a)} and \textbf{(b)} are averaged across $3$ seeds and smoothed using a filter of $500$ bins.}
    \label{fig:average_resource}
\end{figure}

In an ablation analysis, we compare the model in an inference mode for $T = 10{,}000$ time steps and $N_{\max}=100$ boids under $3$ settings, shown in Figure~\ref{fig:average_resource}. In the first setting, the boids have access to both the optimized controller substrate and the learned mutation-operator. In the second setting, the boids have access only to the optimized controller substrate, while the mutation-operator is replaced by additive noise sampled from a uniform distribution, $\sim U(-0.05, 0.05)$. In the third setting, the boids use a randomly initialized controller substrate together with the same noisy mutation process.

In Figure~\ref{fig:average_resource}(a), we plot the average net resource gained per time step by the group, given by
\begin{equation}
    e^{(+)}(t) = \frac{1}{N_{\max}}\sum_{i}\left( e_i^{(g)}(t) - e_i^{(c)}(t) \right).
\end{equation}

It can be seen that the second setting follows the first setting closely, but remains consistently lower prior to saturation. In contrast, the fully ablated setting remains close to zero throughout the rollout. This suggests that most of the baseline performance is carried by the optimized controller substrate, while the learned mutation-operator provides a smaller but systematic improvement.

Further, in Figure~\ref{fig:average_resource}(b), we plot the average positive resources received via the exchange channel by the boids under the same $3$ settings (similar to Equation~\ref{eq:posiive_exchange}). It can be seen that the first two settings maintain positive exchange activity over long time horizons. In contrast, the fully ablated setting remains close to zero.

\section{Discussion and conclusion}

In this work, we found that group-level selection can discover a combination of mutation-operator and controller substrate such that group members occupy multiple resource channels while still undergoing individual-level minimal-criteria selection. Such a feature can help a group mitigate the tragedy-of-the-commons scenario. Rather than choosing a simple segregation of roles across unrelated channels, we opted for an ecologically coupled resource design. In the present model, the same behavioral degrees of freedom simultaneously shape direct gain from grazing, redistribution through local exchange, and loss through metabolic expenditure. In that sense, the model is closer to ecological trade-offs in a common-pool setting than to a simple heterogeneous partitioning of roles~\citep{werner1993ecological, kneitel2004trade, levin2014public}. We also note that exchange use increases over generations even though it is not directly part of the resource fitness term, which is consistent with its indirect contribution to sustaining the colony under turnover.

The role differentiation that emerged was functional and dynamic rather than caste-like and permanent. Boids occupied different roles through their dominant channel usage, and these roles could change over the lifetime of an individual. In that sense, the present system is closer to dynamic task allocation than to fixed caste differentiation~\citep{gordon2016division}. This may be viewed as a precursor to more permanent division of labour, which in evolutionary history likely required additional structural commitment~\citep{simpson2012evolutionary}. Future work can therefore explore model choices that promote more stable role commitment, for example by introducing costs for repeated switching or by modulating channel gain based on recent role occupancy. It will also be interesting to investigate whether such switching is partly mediated by internal state encoded in the CTRNN dynamics, in line with earlier observations of internal-state-modulated collective behaviour in related active-particle systems~\citep{chaturvedi2025emergence}.

A second point of interest is that role proportions become approximately stable near maturity even though individual boids continue to switch roles. It would therefore be interesting to vary the exchange and grazing coefficients to see whether different ecological trade-offs give rise to different steady-state proportions. The aggregation in boid trajectories observed near maturity also deserves closer analysis, especially in relation to the strength of the exchange channel.

Finally, the ablation analysis suggests that most of the baseline performance is carried by the shared controller substrate, while the mutation operator provides a smaller but useful improvement. This leaves open the possibility that alternative mutation-operator designs, longer training horizons, or different hyperparameter regimes may reveal a stronger role for mutation in shaping the collective regime. 

\section{Acknowledgements}
This publication is part of the project Dutch Brain Interface Initiative (DBI$^2$) with project number 024.005.022 of the research programme Gravitation which is (partly) financed by the Dutch Research Council (NWO).

\clearpage

\footnotesize
\bibliographystyle{apalike}
\bibliography{example} 

\end{document}